\documentclass[aip,reprint,amsmath,amssymb,groupedaddress]{revtex4-2}

\draft 

\usepackage{graphicx}
\usepackage{dcolumn}
\usepackage{bm}

\begin{document}


\title{Joule heating and the thermal conductivity of a two-dimensional electron gas at cryogenic temperatures studied by modified 3$\omega$ method} 



\author{Akira Endo}
\email[]{akrendo@issp.u-tokyo.ac.jp}
\affiliation{The Institute for Solid State Physics, The University of Tokyo, 5-1-5 Kashiwanoha, Kashiwa, Chiba 277-8581, Japan}

\author{Shingo Katsumoto}
\affiliation{The Institute for Solid State Physics, The University of Tokyo, 5-1-5 Kashiwanoha, Kashiwa, Chiba 277-8581, Japan}

\author{Yasuhiro Iye}
\affiliation{The Institute for Solid State Physics, The University of Tokyo, 5-1-5 Kashiwanoha, Kashiwa, Chiba 277-8581, Japan}


\date{\today}

\begin{abstract}
During the standard ac lock-in measurement of the resistance of a two-dimensional electron gas (2DEG) applying an ac current $I = \sqrt{2} I_0 \sin(\omega t)$, the electron temperature $T_e$ oscillates with the angular frequency $2 \omega$ due to the Joule heating $\propto I^2$. We have shown that the highest ($T_\mathrm{H}$) and the lowest ($T_\mathrm{L}$) temperatures during a cycle of the oscillations can be deduced, at cryogenic temperatures, exploiting the third-harmonic (3$\omega$) component of the voltage drop generated by the ac current $I$ and employing the amplitude of the Shubnikov-de Haas oscillations as the measure of $T_e$. The temperatures $T_\mathrm{H}$ and $T_\mathrm{L}$ thus obtained allow us to roughly evaluate the thermal conductivity $\kappa_{xx}$ of the 2DEG via the modified 3$\omega$ method, in which the method originally devised for bulk materials is modified to be applicable to a 2DEG embedded in a semiconductor wafer. $\kappa_{xx}$ thus deduced is found to be consistent with the Wiedemann-Franz law. The method provides a convenient way to access $\kappa_{xx}$ using only a standard Hall-bar device and the simple experimental setup for the resistance measurement.
\end{abstract}

\maketitle 

\section{Introduction \label{SecIntr}}
Varieties of techniques have been employed for thermometry required to probe the thermal or thermoelectric properties of a two-dimensional electron gas (2DEG) at cryogenic temperatures. For instance, thermopower across a quantum point contact (QPC) \cite{Appleyard98} or the width of the Coulomb blockade peak at a quantum dot (QD) \cite{Venkatachalam12} has been used to measure the electron temperature $T_e$ of the 2DEG to which the QPC or QD is attached. These techniques require the sophisticated device (QPC or QD) to be fabricated, employing  electron-beam lithography, onto the semiconductor wafer harboring the 2DEG\@.  
A simpler way is to make use of the resistance of the 2DEG itself. In principle, any aspect of the resistance that depends on the temperature can be used as the thermometer. The resistance at the zero \cite{Wennberg86,Syme89,Schmidt12} or a small (non-quantizing) \cite{Mittal96,Mittal96N,Schmidt12JEM} magnetic field, negative magnetoresistance due to the weak localization, \cite{Mittal96,Mittal96N} the amplitude of the Shubnikov-de Haas oscillations (SdHOs), \cite{Hirakawa86,Ma91} the maximum slope of the Hall resistance between two adjacent integer quantum Hall plateaus, \cite{Chow95} activated behavior of a fragile fractional quantum Hall state \cite{Samkharadze11} are among the temperature-dependent phenomena in the resistance that have been used as the thermometer.

For a three-dimensional (3D) bulk material, an interesting ac measurement technique, dubbed the 3$\omega$ method, \cite{Cahill90} is known as a method to measure the thermal conductivity. In this method, resistance of a thin and long metallic film deposited on the surface of a 3D sample to be measured is exploited as both the heater and the thermometer. The temperature oscillations $\Delta T$ generated by an ac heating current with the angular frequency $\omega$ are monitored by detecting  the third-harmonic ($3\omega$) component of the resulting voltage drop (see Sec.\ \ref{SecBasics} for the basic principles). Owing to the cylindrical decay in the 3D sample of $\Delta T$ with the distance $r$ from the wire-like metallic heater, \cite{Carslaw59} the thermal conductivity can be extracted either from the $\ln\omega$ dependence of the real part (in-phase oscillations) or from virtually $\omega$-independent imaginary part (out-of-phase oscillations) of $\Delta T$. \cite{Cahill90}

In the present study, we make an attempt to apply the 3$\omega$ method to a 2DEG at cryogenic temperatures, using a Hall-bar device fabricated from a conventional GaAs/AlGaAs wafer hosting a high-mobility 2DEG\@. The main channel of the Hall-bar device serves simultaneously as the heater and the thermometer. By passing a relatively large ac current $I$ with the angular frequency $\omega$ along the main channel, the electron temperature $T_e$ is raised and oscillates with $2\omega$ due to the Joule heating $\propto I^2$. Note that $T_e$ readily becomes higher than the lattice temperature $T_\mathrm{p}$ of the GaAs crystal hosting the 2DEG due to the weak electron-phonon coupling at low temperatures in this system. \cite{Price82} This \textit{current heating technique} has been widely used for the measurement of the diffusion thermopower, \cite{Gallagher90,Maximov04,Fujita10E} in which the $2\omega$ component representing the voltage induced by the temperature gradient is detected.  In the measurement of the thermopower, however, it suffices to know the time-average of the raised $T_e$. The oscillations of $T_e$ with time have therefore not been paid attention it deserves thus far, on which we shed light in the present paper.  We demonstrate that the highest ($T_\mathrm{H}$) and the lowest ($T_\mathrm{L}$) values of $T_e$ during a cycle of the oscillations can be deduced by measuring the $3\omega$ component of the resistance along the main channel, and then converting the resistance to the temperature via the amplitude of the SdHOs at relatively small magnetic fields. We obtain $T_\mathrm{H}$ and $T_\mathrm{L}$ for various values of the heating current $I$. With these temperatures, we further deduce the thermal conductivity of the 2DEG\@. In this procedure, considerable modification from the original 3$\omega$ method is required due to the differing dimensionality and the heat transfer from the 2DEG to the lattice. The thermal conductivity of the 2DEG is obtained at a fixed $\omega$ from the thermal flux flowing from the main channel to the electrical contacts through the 2DEG in the voltage arms. The high Hall angle approaching $\pi/2$, achieved with a small magnetic field in a high-mobility 2DEG, substantially simplifies the calculation of the thermal flux.
The modified $3\omega$ method described in this paper presents an experimental method to probe the temperature oscillations of a 2DEG due to the Joule heating by an ac current, as well as a convenient way to roughly evaluate the thermal conductivity in the magnetic field, using only a standard Hall-bar device and the experimental setup for the standard ac lock-in resistance measurement.

\section{Basic Principles \label{SecBasics}}
\begin{figure}
\includegraphics[width=8.6cm,clip]{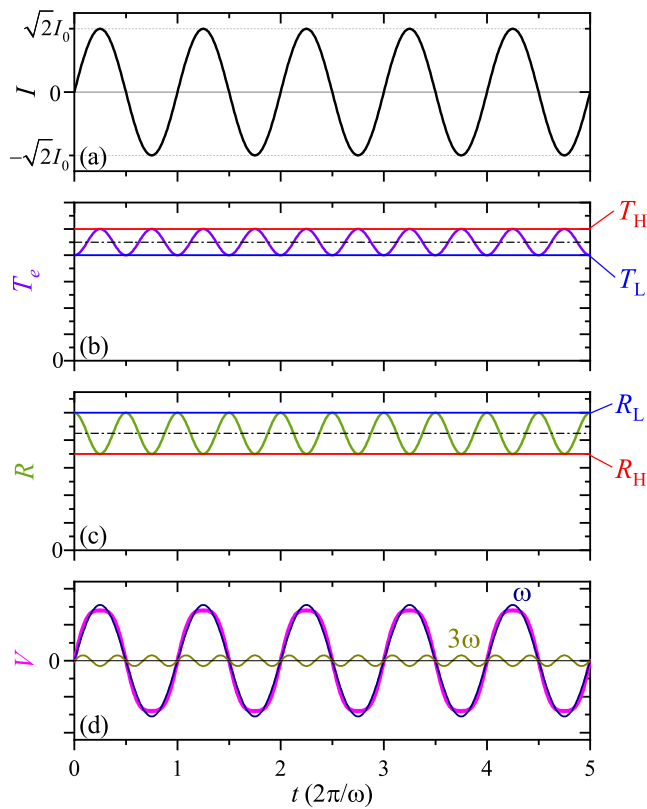}%
\caption{Schematic diagrams depicting the sinusoidal oscillations of (a) the heating current $I$, and the resulting oscillations in (b) the electron temperature $T_e$,  (c) the resistance $R$, and (d) the voltage drop $V = RI$.  $T_\mathrm{L}$ and $T_\mathrm{H}$ represent the lowest and the highest temperatures during a cycle, and $R_\mathrm{L}$ and $R_\mathrm{H}$ are the resistances at $T_\mathrm{L}$ and $T_\mathrm{H}$, respectively. $dR/dT_e <0$ is assumed in this example. The $\omega$ and $3 \omega$ components, $V_\omega$ and $V_{3 \omega}$, contained in $V$ are also plotted with thin lines in (d). \label{timedepschem}}
\end{figure}
We start by outlining the basic principles of monitoring the Joule heating by detecting the third harmonics of the resistance. As in the case in the standard ac lock-in measurement of the resistance, we apply an ac excitation current
\begin{equation}
I = \sqrt{2} I_0\sin{\left(\omega t\right)}, \label{EqI}
\end{equation}
with the angular frequency $\omega$ [Fig.\ \ref{timedepschem}(a)] to the Hall-bar device. To achieve the heating, the rms amplitude $I_0$ of the current is allowed to take a relatively large value (up to several $\mu$A in a Hall-bar device of a GaAs/AlGaAs 2DEG with the width of $\sim$50 $\mu$m). In ordinary resistance measurements at cryogenic temperatures, by contrast, $I_0$ is kept small (typically $I_0 \alt 100$ nA) to avoid the heating. Due to the Joule heating $\propto I^2$, the electron temperature $T_e$ of the 2DEG oscillates with the angular frequency 2$\omega$ between the lowest ($T_\mathrm{L}$) and the highest ($T_\mathrm{H}$) values [Fig.\ \ref{timedepschem}(b)]. If the resistance depends on the temperature, the resistance $R(T_e)$ also oscillates with $2\omega$ between $R_\mathrm{L} \equiv R(T_\mathrm{L})$ and $R_\mathrm{H} \equiv R(T_\mathrm{H})$ [Fig.\ \ref{timedepschem}(c)]. The temperature and the resistance oscillations are approximated well by sinusoidal waves
%
\begin{equation}
T_e =T_\mathrm{L}+\left(T_\mathrm{H}-T_\mathrm{L}\right)\ \frac{1-\cos{\left(2\omega t\right)}}{2} \label{Eqt}
\end{equation} 
and
\begin{equation}
R= R_\mathrm{L}+\left(R_\mathrm{H}-R_\mathrm{L}\right)\ \frac{1-\cos{\left(2\omega t\right)}}{2} \label{EqR}
\end{equation}
respectively, when the relative variations are small, i.e., $T_\mathrm{H}-T_\mathrm{L} \ll T_\mathrm{ave} \equiv (T_\mathrm{H}+T_\mathrm{L})/2$ and $|R_\mathrm{H}-R_\mathrm{L}| \ll R_\mathrm{ave} \equiv (R_\mathrm{H}+R_\mathrm{L})/2$. \footnote{Note that the two approximations are basically independent and thus Eq.\ (\ref{EqR}) remains good approximation as long as $|R_\mathrm{H}-R_\mathrm{L}|$ is small even if $T_\mathrm{H}-T_\mathrm{L}$ is not so small, as is the case in the measurements presented below. Strictly speaking, the oscillations in $R$ affects the Joule heating, which, in turn, alters the oscillations in $R$. The resulting oscillations determined self-consistently inevitably contain higher harmonic terms. However, we neglect these higher-order effects, which are small when $|R_\mathrm{H}-R_\mathrm{L}|$ is small} From Eqs.\ (\ref{EqI}) and (\ref{EqR}), we can see that the voltage drop
\begin{equation}
V=RI= \sqrt{2} V_{\omega} \sin{\left(\omega t\right)} + \sqrt{2} V_{3\omega} \sin{\left(3\omega t\right)} \label{EqV}
\end{equation}
contains the fundamental ($\omega$) and the third-harmonic (3$\omega$) components [Fig.\ \ref{timedepschem}(d)], where
\begin{subequations}
\begin{align}
\frac{V_{\omega}}{I_0} = \frac{3R_\mathrm{H}+R_\mathrm{L}}{4} \equiv R_{\omega} \label{EqR1}, \\
\frac{V_{3\omega}}{I_0} = -\frac{R_\mathrm{H}-R_\mathrm{L}}{4} \equiv R_{3\omega} \label{EqR3}.
\end{align} \label{EqR1R3}
\end{subequations}
Therefore, by picking out the $\omega$ and the 3$\omega$ components $V_{\omega}$ and $V_{3\omega}$ of the (rms) voltage drop employing the lock-in technique, we can deduce the resistance at the lowest and the highest temperatures during the course of the temperature oscillations, given by
\begin{subequations}
\begin{align}
R_\mathrm{L} & = R_{\omega}+3R_{3\omega} \label{EqRL}, \\
R_\mathrm{H} & = R_{\omega}-R_{3\omega} \label{EqRH}.
\end{align} \label{EqRLRH}
\end{subequations}
$R_\mathrm{L}$ and $R_\mathrm{H}$ thus obtained can further be translated to $T_\mathrm{L}$ and $T_\mathrm{H}$ if the temperature dependence $R(T)$ is  known. In the present study, we employ the temperature dependence of the amplitude of the Shubnikov-de Haas oscillations (SdHOs) for the resistance-to-temperature conversion.
As we will show in Sec.\  \ref{SSecThCnd}, the temperature response to the Joule heating thus deduced, combined with the estimation of the power transferred to the lattice of the GaAs substrate hosting the 2DEG, enables us to evaluate the thermal conductivity of the 2DEG\@.

\section{Experimental results and discussion \label{SecExp}}

\subsection{Experimental details \label{SSecExpDtl}}
The Hall-bar device used in the present study was fabricated by standard photo-lithography from a conventional GaAs/AlGaAs wafer containing a 2DEG with the mobility $\mu = 110$ m$^2$/(V s) and the carrier density $n_e = 2.8\times 10^{15}$ m$^{-2}$ just below the heterointerface at the depth of 70 nm from the surface. The measurements were carried out in a dilution refrigerator (Kelvinox TLM, Oxford Instruments), with the Hall-bar device immersed in the mixing chamber of the fridge. The $\omega$ component ($V_\omega$) and the $3\omega$ component ($V_{3 \omega}$) were measured simultaneously using two separate lock-in amplifiers (LI-575, NF Corporation). We performed the measurements with the frequency $f = \omega/(2\pi)$ ranging from 19 to 93 Hz and found that the results are virtually independent of $f$ in this frequency range. \footnote{Since thermalization takes place via the electron-electron and the electron-phonon interactions, and the scattering times for these interactions are of the order of picoseconds \cite{Fukuyama83} and nanoseconds, \cite{Mittal96N,KajiokaCO13}  respectively, in the GaAs-based 2DEG at cryogenic temperatures, the thermal process examined in the present study can be considered to take place instantaneously in the timescale of the measurements. This is confirmed by the absence of the frequency dependence} In what follows, we present the data taken with $f = 73$ Hz at the bath temperature $T_\mathrm{bath} = 15$ mK\@. 
(Data taken at other values of $f$ and $T_\mathrm{bath}$ are presented in the supplementary material.)

\subsection{Dependence of the third harmonics on the heating current \label{SSecDR3I}}
\begin{figure}
\includegraphics[width=8.4cm,clip]{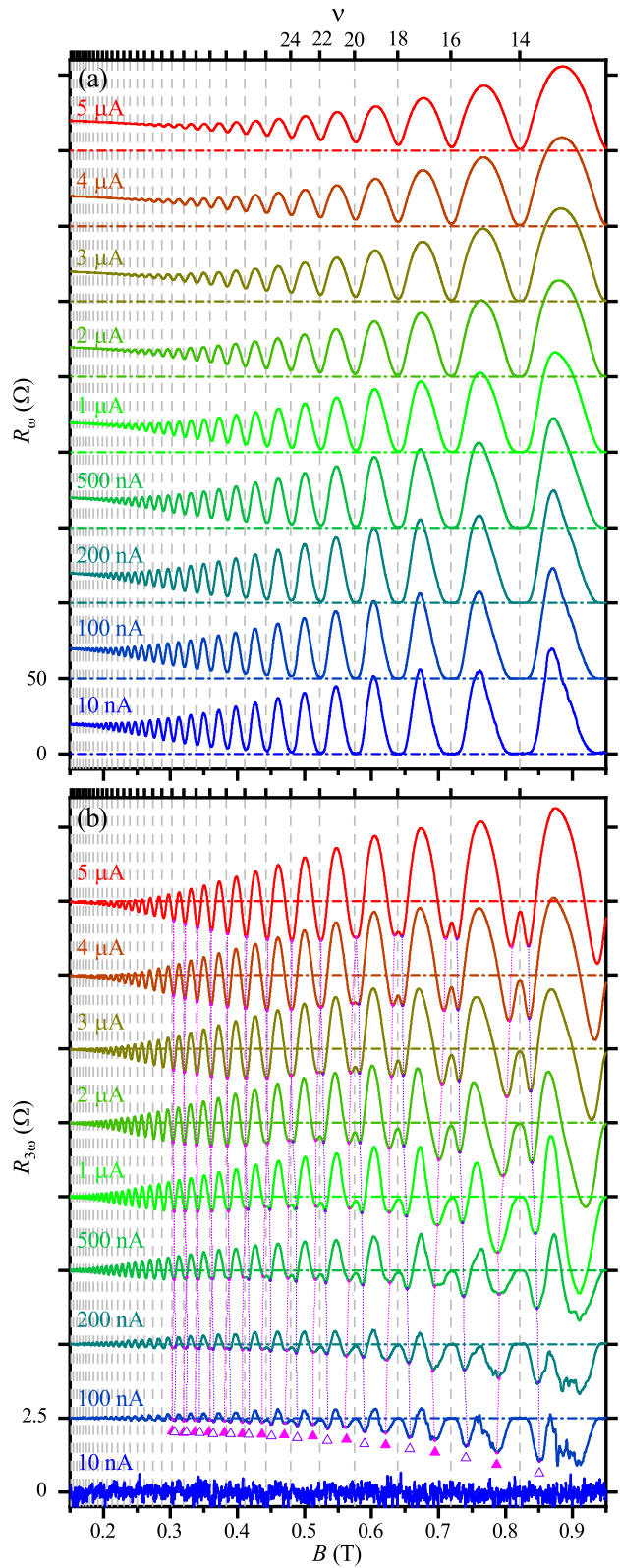}%
\caption{The $\omega$ (a) and $3 \omega$ (b) components of the resistance $R_{\omega}$ and $R_{3\omega}$ for various values of the heating current $I_0$ noted in the figures.  Traces are sequentially offset by 50 and 2.5 $\Omega$ in (a) and (b), respectively, for clarity, with increasing $I_0$. The vertical dashed lines and the horizontal dotted-dashed lines indicate the positions of the even-integer fillings (top axis) and the zero for the corresponding trace, respectively. 
See the text for the upward triangles accompanied by dotted lines in (b).
\label{R1R3Is}}
\end{figure}

In Fig.\ \ref{R1R3Is}, we plot $R_\omega$ and $R_{3\omega}$ given by Eq.\ (\ref{EqR1R3}) obtained from $V_\omega$ and $V_{3\omega}$ measured with the lock-in amplifiers for various values of $I_0$ ranging from 10 nA to 5 $\mu$A\@. $R_\omega$ is nothing but the resistance obtained by ordinary ac lock-in measurement. It exhibits SdHOs which develop, with the increase of the magnetic field $B$, into the quantum Hall effect (QHE) state characterized by the flat areas of $R_\omega = 0$ having a finite magnetic-field span [Fig.\ \ref{R1R3Is}(a)]. Only spin-unresolved even-integer QHE states are observed in the magnetic-field range shown in the figure. [The Landau-level filling factor $\nu = n_e h / (eB)$ is shown by the top axis.]
Following the increase in $I_0$, the amplitude of SdHOs diminishes and the flat areas of QHE shrinks toward the center. As is well known, these behaviors are attributable to the raised electron temperature due to the Joule heating. Our interest in the present study is mainly on the behavior of $R_{3\omega}$ representing the decrement (multiplied by $1/4$) of the resistance, while the temperature varies from the lowest ($T_\mathrm{L}$) to the highest ($T_\mathrm{H}$) values within a cycle [Eq.\ (\ref{EqR3})]. As can be seen in Fig.\ \ref{R1R3Is}(b), the Joule heating is apparently not enough to generate the temperature oscillations detectable with $R_{3\omega}$ at $I_0 = 10$ nA. Oscillations in $R_{3\omega}$ become apparent at $I_0 = 100$ nA and the amplitude initially increases with $I_0$ reflecting the increase in $\Delta T = T_\mathrm{H} - T_\mathrm{L}$. With further increase in $I_0$, however, the amplitude starts to decline, gradually from the lower $B$ side. \footnote{Close inspection of the oscillation amplitudes reveals that the declining commences above $I_0 = 1$, 2, 3, and 4 $\mu$A in the magnetic-field region $B < $ 0.24 T, 0.24 T $< B <$ 0.31 T, 0.31 T $< B <$ 0.36 T, and 0.36 T $< B <$ 0.40 T, respectively} This is because the effect of increasing $\Delta T$ becomes overridden by the decrease in the amplitude of SdHOs with the increase in $T_\mathrm{ave}$, and the declination is more prominent for a lower $B$.  At low $B$, $R_{3\omega}$ exhibits oscillations in phase with those of $R_{\omega}$. This indicates that the decrement in the resistance by $\Delta T$ takes a local maximum and a local minimum at the peaks and the troughs of the resistance, respectively, resulting in the diminished amplitude of the SdHOs. The locations of the peaks remain the same between $R_{3\omega}$ and $R_{\omega}$ up to higher $B$, consistent with the decreasing resistance with $\Delta T$ at the peaks. In the vicinity of minima in $R_\omega$, by contrast, $R_{3\omega}$ takes on rather complicated line shape at higher $B$. The flat areas of $R_{3\omega} = 0$ appear at the locations corresponding to $R_\omega = 0$ areas at high-$B$ and low-$I_0$ regions, where the QHE is robust and hardly affected by $\Delta T$. Two minima flanking a flat area are seen, which represent the increase in the resistance caused by $\Delta T$ in the region adjacent to the flat area leading to the shrinking of the flat area. The minima on the higher-field (lower-field) side of a flat area are indicated by open (solid) upward triangles for $I_0 = 100$ nA, and the accompanying dotted lines are eye-guides to follow the shift of the minima with increasing $I_0$.  With the decrease of $B$ and/or increase of $I_0$, the two minima get close to each other narrowing the flat area, thereby generating the line shape containing alternating high and low peaks. The two minima eventually merge and engulf the flat region, signaling the transition from QHE to SdHOs. Note that the peculiar line shape in $R_{3\omega}$ with alternating peak heights survives down to the $B$-range where QHE is not apparent in $R_\omega$ ($0.3 \alt B \alt 0.5$ T for $I_0 = 100$ nA and 200 nA), revealing the presence of the precursor of the QHE state in these regions.

\subsection{Response of the electron temperature to the Joule heating \label{SSecTe}}
\begin{figure*}
\includegraphics[width=17.2cm,clip]{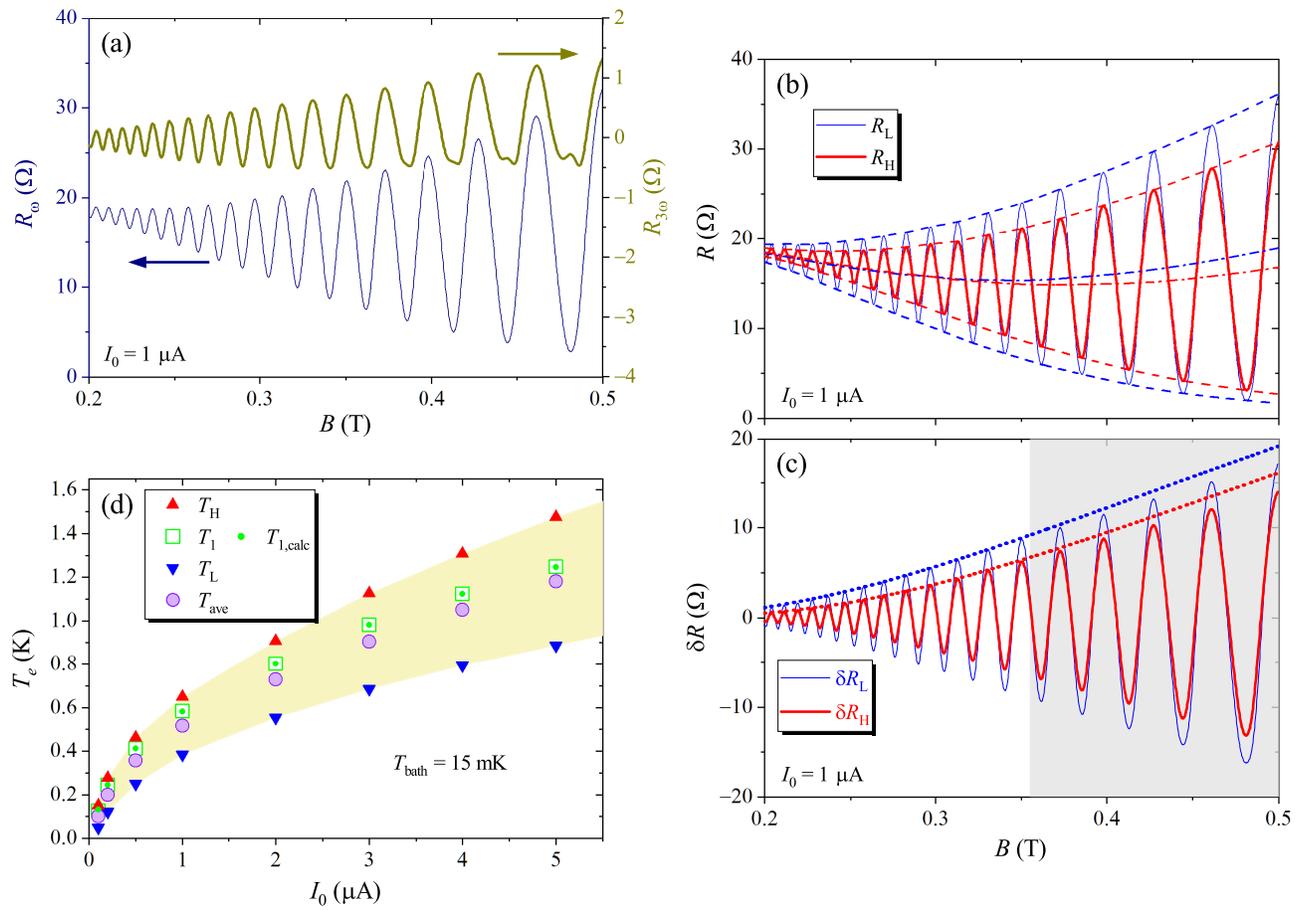}%
\caption{(a)-(c) Illustration of the derivation of $T_\mathrm{L}$ and $T_\mathrm{H}$ from $R_{\omega}$ and $R_{3 \omega}$, taking the case for $I_0 =1$ $\mu$A as an example. (a) $R_{\omega}$ (thin line, left axis) and $R_{3 \omega}$ (thick line, right axis). (b) $R_\mathrm{L}$ (thin solid line) and $R_\mathrm{H}$ (thick solid line) calculated by Eq.\ (\ref{EqRLRH}) from $R_{\omega}$ and $R_{3 \omega}$ in (a). Dashed lines represent the upper ($R_\mathrm{up}$) and lower ($R_\mathrm{lw}$) envelopes of the SdHOs and the dotted-dashed lines are their averages $R_\mathrm{bg} = (R_\mathrm{up}+R_\mathrm{lw})/2$, which are taken as the slowly-varying backgrounds. (c) The oscillatory parts $\delta R_\mathrm{L}$ (thin solid line) and $\delta R_\mathrm{H}$ (thick solid line) obtained by subtracting the corresponding background $R_\mathrm{bg}$ from $R_\mathrm{L}$ and $R_\mathrm{H}$ in (b). The dotted lines represent Eq.\ (\ref{EqSdHamp}) with $T_e = T_\mathrm{L} = 0.39$ K and $T_e = T_\mathrm{H} = 0.65$ K, which reproduce the amplitude of $\delta R_\mathrm{L}$ and $\delta R_\mathrm{H}$, respectively, below $\sim$0.35 T\@. Shaded area indicates the range of the magnetic field where the amplitudes are reduced due to the effect of incipient spin-splitting. (d) $T_\mathrm{L}$ and $T_\mathrm{H}$ for various values of $I_0$, obtained by following the procedure exemplified in (a)-(c). The average temperature, $T_\mathrm{ave} = (T_\mathrm{L}+T_\mathrm{H})/2$, and the temperature obtained from the analysis of the SdHOs in $R_\omega$, $T_1$, are also plotted.  $T_1$ shows good agreement with $T_{1,\mathrm{calc}}$ (indicated by the small dots) calculated by Eq.\ (\ref{T1calc}) from $T_\mathrm{L}$ and $T_\mathrm{H}$. \label{TLTHderiv}}
\end{figure*}

In this section, we deduce the lowest ($T_\mathrm{L}$) and the highest ($T_\mathrm{H}$) temperatures for various $I_0$ from the measured $R_{\omega}$ and $R_{3 \omega}$ shown in Sec.\  \ref{SSecDR3I}. The procedure is illustrated in Fig.\ \ref{TLTHderiv}, taking the case for $I_0 = 1$ $\mu$A as an example. The first step is to transform $R_{\omega}$ and $R_{3 \omega}$ [Fig.\ \ref{TLTHderiv}(a)] into $R_\mathrm{L}$ and $R_\mathrm{H}$, the resistances of the 2DEG device at the moments when the electron temperatures are $T_\mathrm{L}$ and $T_\mathrm{H}$, respectively, using Eq.\ (\ref{EqRLRH}). The resulting $R_\mathrm{L}$ and $R_\mathrm{H}$ are shown in Fig.\ \ref{TLTHderiv}(b). They exhibit SdHOs with differing amplitudes, reflecting the difference in the temperature. 

The next step is to deduce $T_\mathrm{L}$ and $T_\mathrm{H}$ from $R_\mathrm{L}$ and $R_\mathrm{H}$.
To this end, we employ the well-known temperature and magnetic-field dependence of the amplitude $\delta R_\mathrm{amp}$ of the SdHOs \cite{Coleridge91},
\begin{equation}
\frac{\delta R_\mathrm{amp}}{4R_0} = \exp{\left( -\frac{\pi}{\mu_\mathrm{q} B} \right)} A\left( \frac{T_e}{T_\mathrm{c}} \right), \label{EqSdHamp}
\end{equation}
for the resistance-to-temperature conversion, where
$A(x) \equiv x / \sinh{x}$, $T_\mathrm{c} \equiv \hbar \omega_\mathrm{c} / (2\pi^2 k_\mathrm{B})$ with $\omega_\mathrm{c} = eB/m^*$ being the cyclotron angular frequency and $m^*$ being the effective mass, $R_0$ is the resistance at $B = 0$, and $\mu_\mathrm{q}$ represents the quantum mobility. The first and the second factors represent the damping of the amplitude by the impurity scattering and the temperature, respectively.
In order to apply Eq.\ (\ref{EqSdHamp}), we need to extract the amplitude of the SdHOs from $R_\mathrm{L}$ and $R_\mathrm{H}$. This is done by the method we used to extract the oscillatory component detailed in a previous publication \cite{Endo00e}: briefly, upper and lower envelope curves $R_\mathrm{up}$ and $R_\mathrm{lw}$ are defined as spline curves smoothly connecting the maxima and minima, respectively, and with them we obtain the slowly-varying background $R_\mathrm{bg} = (R_\mathrm{up} + R_\mathrm{lw})/2$, the oscillatory part $\delta R = R - R_\mathrm{bg}$, and the amplitude $\delta R_\mathrm{amp} = (R_\mathrm{up} - R_\mathrm{lw})/2$. The oscillatory parts $\delta R_\mathrm{L}$ and $\delta R_\mathrm{H}$ contained in $R_\mathrm{L}$ and $R_\mathrm{H}$, respectively, extracted by this procedure are plotted in Fig.\ \ref{TLTHderiv}(c).

The value of $\mu_\mathrm{q}$ in Eq.\ (\ref{EqSdHamp}) is determined from $R_\omega$ at $I_0 = 10$ nA\@. [Details of the procedure is presented in the supplementary material IV.] At this current, the Joule heating is negligibly small [$R_{3\omega} \simeq 0$, see Fig.\ \ref{R1R3Is}(b), and thus $R_\mathrm{L} \simeq R_\mathrm{H} \simeq R_\omega$] and $T_e$ is expected to be close to the bath temperature $T_\mathrm{bath} = 15$ mK\@. \footnote{Since the sample, as well as the wires connected to the sample (several centimeters long at the connected end), is immersed in the mixture of the dilution fridge, and the current source is appropriately filtered before being connected to the wires, we believe that $T_e$ is not substantially higher than $T_\mathrm{bath}$. We also have ample experimental evidence, including the developing fractional quantum Hall states or SdH amplitudes, showing that $T_e$ keeps going down while $T_\mathrm{bath}$ is cooled down to the base temperature. Note, however, that the following analysis is valid even if $T_e$ is heated to slightly above $T_\mathrm{bath}$, so long as $T_e \ll T_\mathrm{c}$ and thus $A(T_e/T_\mathrm{c}) \simeq 1$. The value of $\mu_\mathrm{q}$ obtained by replacing $A(T_e/T_\mathrm{c})$ with 1 is virtually indistinguishable from that deduced in the main text}
We extract $\delta R_\mathrm{amp}$ from $R_\omega$ following the procedure described above. Substituting $T_e = T_\mathrm{bath}$  and using $\mu_\mathrm{q}$ as a fitting parameter, we search for the value of $\mu_\mathrm{q}$ with which Eq.\ (\ref{EqSdHamp}) reproduces the $B$-dependence of $\delta R_\mathrm{amp}$. [This is basically the same as the well-known method using the Dingle plot. \cite{Coleridge91}] We find that excellent agreement is achieved for $B \alt 0.35$ T with $\mu_\mathrm{q} = 4.1$ m$^2$/(V s). Slight deviation at higher magnetic fields [Fig.\ S6 (b) in the supplementary material] is attributable to the onset of the spin-splitting. Although spin-resolved odd-integer QHE states are not observed in the magnetic-field range examined in the present paper as mentioned in Sec.\  \ref{SSecDR3I}, the commencement of incomplete spin splitting is known to reduce the peak height of the SdHOs, \cite{Endo08SdHH} leading to the deviation from Eq.\ (\ref{EqSdHamp}). In what follows, therefore, we focus on the low magnetic-field range ($B \alt 0.35$ T), where SdHOs remain virtually unaffected by the spin splitting. In the range of $T_e$ encompassed in the present study, $\mu_\mathrm{q}$ is independent of the temperature, since the mobility is limited predominantly by the impurity scattering and the contribution from the electron-phonon scattering is negligibly small. \cite{Walukiewicz84,Davies98B} This allows us to use the same value of $\mu_\mathrm{q}$ deduced here in the analysis of the data taken at higher $I_0$.

Substituting $\mu_\mathrm{q} = 4.1$ m$^2$/(V s) and using $T_e$ as the fitting parameter, the fitting of Eq.\ (\ref{EqSdHamp}) to the amplitudes of the SdHOs extracted from $R_\mathrm{L}$ and $R_\mathrm{H}$  allows us to deduce the electron temperatures $T_\mathrm{L}$ and $T_\mathrm{H}$, as exemplified in Fig.\ \ref{TLTHderiv}(c). As mentioned above, the fitting is performed in the limited magnetic field range ($B \alt 0.35$ T). Here again, we can see slight deviation caused by the incipient incomplete spin splitting at higher magnetic fields [the shaded region in Fig.\ \ref{TLTHderiv}(c)].

In  Fig.\ \ref{TLTHderiv}(d), we compile $T_\mathrm{L}$ and $T_\mathrm{H}$ obtained for various values of $I_0$ ranging from 100 nA to 5 $\mu$A\@. The figure, representing one of the highlights in the present study, shows how the average temperature $T_\mathrm{ave} = (T_\mathrm{L} + T_\mathrm{H})/2$ and the temperature increment $\Delta T = T_\mathrm{H} - T_\mathrm{L}$ increase with $I_0$.
A more conventional way to estimate the electron temperature of a 2DEG heated by a current is to perform the analysis of the amplitude of SdHOs directly on $R_\omega$. \cite{Hirakawa86,Fletcher92,Fujita10E} 
We have applied the same analysis described above to the SdHOs in $R_\omega$. The resulting  electron temperature $T_1$, also plotted in Fig.\ \ref{TLTHderiv}(d), reveals that the conventional method yields temperatures slightly higher than $T_\mathrm{ave}$. For more quantitative account of $T_1$, we recall the relation between $R_\omega$ and $R_\mathrm{L}$, $R_\mathrm{H}$,  Eq.\ (\ref{EqR1R3}), which leads to the relation between the temperatures,
\begin{equation}
A\left( \frac{T_{1}}{T_\mathrm{c}} \right) = \frac{3A(T_\mathrm{H}/T_\mathrm{c})+A(T_\mathrm{L}/T_\mathrm{c})}{4}. \label{T1calc}
\end{equation}
In Fig.\ \ref{TLTHderiv}(d), we also plot values of $T_1$ numerically calculated with Eq.\ (\ref{T1calc}) from $T_\mathrm{L}$ and $T_\mathrm{H}$, which show excellent agreement with $T_1$ obtained from the direct analysis of $R_\omega$.

\subsection{Thermal conductivity of a 2DEG \label{SSecThCnd}}
\begin{figure}
\includegraphics[width=8.6cm,clip]{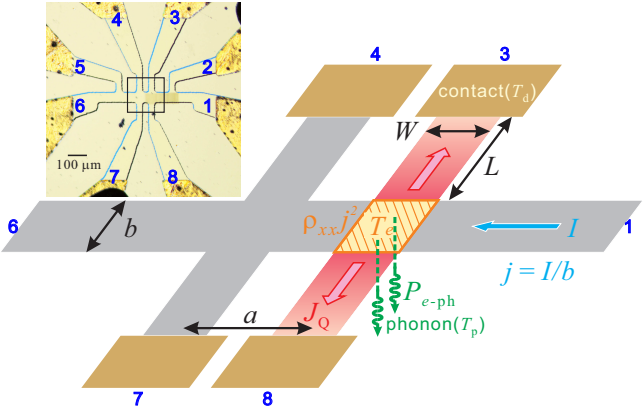}%
\caption{Schematic illustration of the thermal flow in a Hall bar device. $\rho_{xx} j^2$: Joule heating per area by the current density $j = I/b$. $J_\mathrm{Q}$: thermal flux due to diffusion into the contact pads.  $P_{e\textrm{-ph}}$: power per area transferred to the lattice via the electron-phonon interaction. Inset: Optical micrograph of the Hall bar device used in the present study. The main panel corresponds to the area enclosed by the rectangle. \label{HallBar}}
\end{figure}

The response of the temperature to the Joule heating drawn out in Sec.\  \ref{SSecTe} allows us to roughly estimate the thermal conductivity $\kappa_{xx}$ of the 2DEG\@. We consider a simple Hall-bar device having the main channel with the width $b$ and the voltage arms with the width $W$ and the length $L$, as schematically depicted in Fig.\ \ref{HallBar}. The current $I$ passing through the main channel introduces Joule heat $\rho_{xx} j^2$ per area to the 2DEG residing in the main channel, where $\rho_{xx}$ is the resistivity of the 2DEG and $j = I/b$ is the current density. Now we focus on the hatched area in Fig.\ \ref{HallBar}, namely, the part of the main-channel area in direct connection to the voltage arms. The heat deposited to this area by the Joule heating, $\rho_{xx} j^2 b W$, is lost by diffusion through the voltage arms into the electric contact ($J_\mathrm{Q}$) or by being imparted to the lattice via the electron-phonon interaction ($P_{e\textrm{-ph}}$). \footnote{The contribution of the heat capacity is totally negligible due to the extremely small electronic specific heat $C_e$ of a 2DEG, \cite{Zawadzki84} $C_e / T \sim 1 \times 10^{-9}$ J/(m$^2$K$^2$) at $B \alt 0.3$ T, and thus $\omega \int_{T_\mathrm{L}}^{T_\mathrm{H}} C_e dT \ll \rho_{xx} j^2$, $J_\mathrm{Q}/(bW)$, $P_{e\textrm{-ph}}$}

In a recent publication, \cite{Endo19} the present authors have shown that the thermal flux from the high-temperature end ($T_\mathrm{s}$) to the low-temperature end ($T_\mathrm{d}$) along the length of the rectangular 2DEG area (length $L$ and width $W$), when placed in a magnetic field, is given by
\begin{equation}
J_\mathrm{Q}(T_\mathrm{s},T_\mathrm{d}) = \frac{K_{xx}}{\cos{\delta}}\frac{\alpha}{S\left(\delta,\alpha\right)}\frac{{T_\mathrm{s}}^2-{T_\mathrm{d}}^2}{2}, \label{EqDiffusion}
\end{equation}
where $\alpha = W/L$ is the aspect ratio, $\delta = \arctan(\sigma_{yx}/\sigma_{xx})$ is the Hall angle with $\sigma_{yx}$ and $\sigma_{xx}$ the Hall conductivity and the diagonal conductivity, respectively, 
\begin{eqnarray}
S\left( {\delta ,\alpha } \right) \equiv \hspace{70mm} \nonumber \\ 
 \int_0^1 \!\!\! {\cos \left\{ {4\delta \sum\limits_{n = 1}^\infty  {\frac{{\sin \left[ {\left( {2n - 1} \right)\pi \xi } \right]}}{{\left( {2n - 1} \right)\pi }}} {\mathop{\rm sech}\nolimits} \left[ {\left( {2n - 1} \right)\frac{{\alpha \pi }}{2}} \right]} \right\} d\xi }, \nonumber \\ \label{Iinteg}
\end{eqnarray}
and
\begin{equation}
K_{xx} \equiv \kappa_{xx}/T_e. \label{Kxx}
\end{equation}
In the derivation of Eq.\ (\ref{EqDiffusion}), we assumed that $K_{xx}$ is independent of $T_e$, which is good approximation for 2DEGs at low temperatures ($T_e \alt 1$ K). \cite{Endo19} 
We have also shown that $S(\delta,\alpha) \simeq \alpha$ for $\alpha \alt 0.5$ in a moderate-to-high magnetic field where $\delta$ approaches $\pi/2$ in a high-mobility 2DEG, and thus $J_\mathrm{Q}(T_\mathrm{s},T_\mathrm{d})$ becomes independent of $\alpha$,
\begin{equation}
J_\mathrm{Q}(T_\mathrm{s},T_\mathrm{d}) \simeq \frac{K_{xx}}{\cos{\delta}}\frac{{T_\mathrm{s}}^2-{T_\mathrm{d}}^2}{2}. \label{EqDiffusionA}
\end{equation}
This allows us to apply Eq.\ (\ref{EqDiffusionA}), to good approximation, to a voltage arm of a Hall-bar device even if the arm is not of rectangular shape, so long as the arm is long enough. The Hall-bar device used in the present study is shown in the inset of Fig.\ \ref{HallBar}. Although the arms retain rectangular shape only for a certain distance from the main channel, the small aspect ratio $\alpha$ throughout the arms  justifies, to a certain degree, the use of the approximation Eq.\ (\ref{EqDiffusionA}).

The power per area transferred from the 2DEG at the temperature $T_e$ to the phonons at the temperature $T_\mathrm{p}$ is written as \cite{Endo19,Price82,KajiokaCO13}
\begin{equation}
P_{e\textrm{-ph}}(T_e,T_\mathrm{p}) = P_l^\mathrm{df}(T_e,T_\mathrm{p}) + P_l^\mathrm{pz}(T_e,T_\mathrm{p}) + 2P_t^\mathrm{pz}(T_e,T_\mathrm{p}), 
\label{EqPhonon}
\end{equation}
with
\begin{equation}
P_s^r(T_e,T_\mathrm{p}) = \Pi_s^r(T_e) - \Pi_s^r(T_\mathrm{p}), \label{EqPi}
\end{equation}
where the deformation-potential coupling and the piezoelectric coupling are denoted by $r$ = df and pz, respectively, and $s = l$ and $t$ represent the longitudinal and the transverse modes, respectively. The functions \cite{Price82,KajiokaCO13} $\Pi_s^r(T)$ are given by \footnote{Strictly speaking, these formulas are for $B = 0$. In the low magnetic-field range we employed for the analysis, however, the Landau quantization  only superposes small ripples on the energy-independent density of states (DOS) at $B = 0$, rather than transforming the DOS into separated discrete levels.  We therefore consider these formulas to serve as good approximation}
\begin{subequations}
\begin{align}
\Pi_l^\mathrm{df}(T) = \frac{e^2D^2v_l{a_\mathrm{B}^\ast}^2}{{64\sqrt2\pi^{\frac{5}{2}}\mu_\mathrm{B}^\ast}^2\rho}\left(\frac{k_\mathrm{B} T}{\hbar v_l}\right)^7\frac{G_l^{\mathrm{df}}(n_e,T)
}{\sqrt{n_e}}  \label{EqPidf} \\
\Pi_s^\mathrm{pz}(T) = \frac{e^4 {h_{14}}^2v_s{a_\mathrm{B}^\ast}^2}{{64\sqrt2\pi^{\frac{5}{2}}\mu_\mathrm{B}^\ast}^2\rho}\left(\frac{k_\mathrm{B} T}{\hbar v_s}\right)^5\frac{G_s^{\mathrm{pz}}(n_e,T)
}{\sqrt{n_e}} \label{EqPipz} \\
(s = l, t). \nonumber
\end{align}
\label{EqPi}
\end{subequations}
The material parameters for GaAs in Eq.\ (\ref{EqPi}) used in the calculations below are as follows: the deformation potential $D = -8.33$ eV, \cite{VdWalle89}  the piezoelectric constant $h_{14} = 1.2\times10^9$ V/m, \cite{Lee83,Lyo88,Lyo89} the longitudinal sound velocity $v_l = 5.14\times 10^3$ m/s, \cite{Lyo88,Lyo89} the transverse sound velocity $v_t = 3.04 \times 10^3$ m/s, \cite{Lyo88,Lyo89} the mass density $\rho = 5.3$ g/cm$^3$, \cite{Blakemore82} the effective Bohr radius $a_\mathrm{B}^* = 10.4$ nm, \cite{KajiokaCO13} and $\mu_\mathrm{B}^* \equiv e\hbar/(2 m^*) = 0.864$ meV/T with the effective mass $m^* = 0.067 m_e$, \cite{Adachi85} where $m_e$ is the bare electron mass. The dimensionless functions \cite{Price82,KajiokaCO13} $G_s^r(n_e,T)$ are detailed in the Appendix.

By balancing the incoming and outgoing thermal flux at the hatched area in Fig.\ \ref{HallBar}, we have 
\begin{equation}
\rho_{xx} j^2  bW=2  J_\mathrm{Q}(T_e,T_\mathrm{d})+P_{e\textrm{-ph}}(T_e,T_\mathrm{p})  bW, \label{EqBalance}
\end{equation}
from which we arrive at the expression for the thermal conductivity 
\begin{equation}
\kappa_{xx} = K_{xx} T_e  = bW\frac{T_e}{{T_e}^2-{T_\mathrm{d}}^2}\cos{\delta}\left[\rho_{xx} j^2-P_{e\textrm{-ph}}\left(T_e,T_\mathrm{p}\right)\right]. \label{Eqthermalcond}
\end{equation}
To proceed further, it is necessary to make several assumptions regarding the temperatures to be substituted into Eq.\ (\ref{Eqthermalcond}).  As can be seen in Fig.\ \ref{timedepschem}, $T_\mathrm{L}$ corresponds to the electron temperature of the 2DEG at the moment the Joule heating vanishes. Therefore, we consider that $T_\mathrm{L}$ also represents the lattice temperature of the part of the substrate in direct contact with the 2DEG, namely, $T_\mathrm{p} = T_\mathrm{L}$; the thin layer buried in the substrate, located in the vicinity of GaAs/AlGaAs heterojunction and hosting the 2DEG having the thickness of the wavefunction $\sim$10 nm, can be heated,  via the electron-phonon interaction, to a lattice temperature higher compared to the rest of the substrate. On the other hand, we assume that the temperature of the electrical contact is the same as the bath temperature, $T_\mathrm{d} = T_\mathrm{bath}$, since the contact pad made of a thin metallic (AuGeNi) film has a granular surface (see the inset to Fig.\ \ref{HallBar}) and therefore expected to be in good thermal contact with the surrounding helium liquid, \cite{Endo19}  in marked contrast with the thin slab buried in the GaAs substrate. Taking the average temperature $T_\mathrm{ave}$ as resulting from the root mean square value of the Joule heating, we substitute $T_e = T_\mathrm{ave}$ and $\rho_{xx} j^2 = R_\mathrm{bg,ave}  I_0^2 /(ab)$ into Eq.\ (\ref{Eqthermalcond}), where we used $R_\mathrm{bg,ave} = (R_\mathrm{L,bg} + R_\mathrm{H,bg})/2$, the average of the backgrounds of $R_\mathrm{L}$ and $R_\mathrm{H}$ [exemplified by the dotted-dashed lines in Fig.\ \ref{TLTHderiv} (b)], \footnote{Note that $R_\mathrm{L,bg}$ and $R_\mathrm{H,bg}$ are almost the same in the area unaffected by the spin splitting. Their separation at higher magnetic fields is attributable to the differing effect of the spin splitting at different temperatures} to evaluate the resistivity $\rho_{xx} = R_\mathrm{bg,ave} b/a$ responsible for the Joule heating, \footnote{Noting that the amplitude of SdHOs is relatively small compared to the background in the magnetic-field range considered here, we neglected the variation of the Joule heating caused by SdHOs. Enhanced and reduced Joule heating at the maxima and the minima of the SdHOs, respectively, lead to the reduction and the enhancement of the amplitude. Both result in the downward shift of the resistance and therefore their net effect on the amplitude is, more or less, expected to be cancelled} with $a$ representing the distance between the voltage arms employed for the measurement of $R_{xx}$ (see Fig.\ \ref{HallBar}).
We also used the same $\rho_{xx}$ along with the semiclassical Hall resistivity \footnote{Experimentally measured Hall resistivity exhibits the onset of quantum Hall plateaus, but the deviation from the semiclassical Hall resistivity is negligibly small in the magnetic-field range considered here} $\rho_{xy} = B/(n_ee)$ to calculate the Hall angle $\delta = \arctan(\rho_{xy}/\rho_{xx})$.
The dimensions of the Hall-bar device in the present study are $a = 62$ $\mu$m, $b = 52$  $\mu$m, and $W = 32$  $\mu$m.

\begin{figure}
\includegraphics[width=8.6cm,clip]{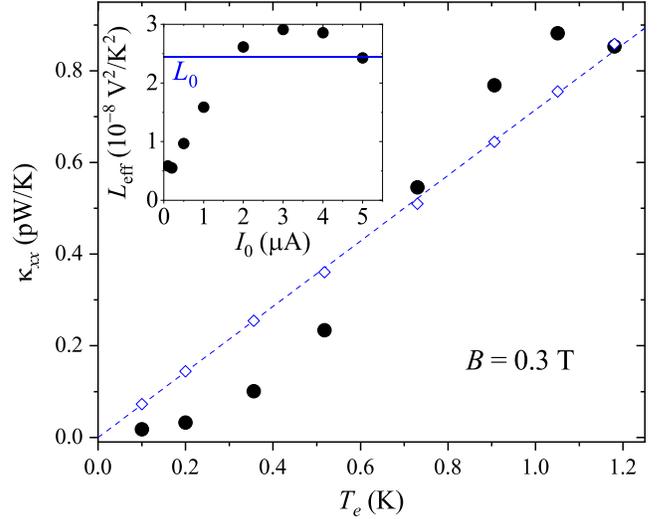}%
\caption{Thermal conductivity $\kappa_{xx}$ calculated by Eq.\ (\ref{Eqthermalcond}) with $T_e = T_\mathrm{ave}$ at $B = 0.3$ T (solid circles). $\kappa_{xx}^\mathrm{WF}$ calculated by the Wiedemann-Franz law, Eq.\ (\ref{EqWF}), is plotted by open diamonds. (Dashed line is an eye guide). Inset: effective Lorenz number $L_\mathrm{eff}$ given by Eq.\ (\ref{Lorenz}) plotted against $I_0$. The horizontal line indicates the value $L_0$ corresponding to the Wiedemann-Franz law. \label{thermalcond}}
\end{figure}

In the main panel of Fig.\ \ref{thermalcond}, we plot $\kappa_{xx}$ thus calculated for various values of $I_0$ at $B = 0.3$ T as a function of $T_e = T_\mathrm{ave}$. 
We also plot, for comparison, the thermal conductivity calculated from the electrical conductivity $\sigma_{xx} = \rho_{xx}/(\rho_{xx}^2+\rho_{xy}^2)$ employing the Wiedemann-Franz law, valid at low temperatures, \footnote{The Wiedemann-Franz (WF) law is expected to be valid at $T \ll E_\mathrm{F}/k_\mathrm{B}$ for systems following the Boltzmann transport equations. Strictly speaking, the rapidly-varying oscillatory part of the SdH oscillations further requires $T \ll T_\mathrm{c}$. In the present discussion, however, we consider the WF law only for the slowly-varying background. Unfortunately, we are unaware of experimental evidence for the WF law to be valid for GaAs-based 2DEGs in the temperature and the magnetic-field range considered in the present study. At $B = 0$ and $2K \alt T \alt 6K$, however, the WF law was experimentally confirmed in Ref.\ \onlinecite{Syme89}}
\begin{equation}
\kappa_{xx}^\mathrm{WF} = L_0 T_e \sigma_{xx}, \label{EqWF}
\end{equation}
where $L_0 = \pi^2 {k_\mathrm{B}}^2 / (3 e^2) = 2.44\times10^{-8}$ V$^2$/K$^2$ is the Lorenz number.
Slight deviation of the  $\kappa_{xx}^\mathrm{WF}$ from the simple linear-$T$ behavior (indicated by the dashed line) is due to the small variation of $\sigma_{xx}$ with $I_0$. We can see that $\kappa_{xx}$ obtained by Eq.\ (\ref{Eqthermalcond}) roughly follows the increasing trend of $\kappa_{xx}^\mathrm{WF}$ with $T_e$.
Comparison with the Wiedemann-Franz law can be made more clearly in the inset to Fig.\ \ref{thermalcond}, in which we plot the effective Lorenz number
\begin{equation}
L_\mathrm{eff} \equiv \frac{\kappa_{xx}}{T_e \sigma_{xx}} = \frac{K_{xx}}{\sigma_{xx}} \label{Lorenz}
\end{equation}
against $I_0$. Although $L_\mathrm{eff}$ coincides with the Wiedemann-Franz value $L_0$ within an order of magnitude, deviations are apparent especially at low $I_0$ (low $T_e$) region. We consider that the low-$T_e$ deviation mainly results from the assumption $T_\mathrm{d} = T_\mathrm{bath}$ mentioned above, suggesting the possibility that the electric contacts are heated above the temperature of the surrounding helium bath by the inflowing thermal flux. 
Slight increase in $T_\mathrm{d}$ let the $\kappa_{xx}$ approach $\kappa_{xx}^\mathrm{WF}$ by diminishing the denominator in Eq.\ (\ref{Eqthermalcond}) when $T_e$ is low and relatively close to $T_\mathrm{d}$.
With the increase of $T_e$, the role of the term $P_{e\textrm{-ph}}(T_e, T_\mathrm{p})$ in Eq.\ (\ref{Eqthermalcond}) becomes acceleratingly important, since it contains the terms varying as ${T_e}^5$ or ${T_e}^7$. At higher $T_e$, therefore, the accuracy of the deduced $\kappa_{xx}$ is severely limited by the preciseness of the values of $D$ and $h_{14}$, which exhibit relatively wide variations among the literature \cite{Blakemore82,Adachi85,Wolfle70,Lee83,Price84,Hirakawa86,Lyo88,Lyo89,VdWalle89,Cakan16} ($|D| = 6.7$--$11$ eV, $h_{14} = 1.2$--$1.45\times10^9$ V/m). We have chosen the values of $D$ and $h_{14}$ noted above, rather arbitrarily, to achieve the fairly good agreement between $\kappa_{xx}$ and $\kappa_{xx}^\mathrm{WF}$ shown in Fig.\ \ref{thermalcond}. 
Considering that a rather simplistic model, represented by Fig.\ \ref{HallBar} and Eq.\ (\ref{EqDiffusionA}), \footnote{We have neglected the heat transferred to the phonon from the arms for simplicity. The possible effect of non-rectangular shape of the arms, potentially leading to the deviation from Eq.\ (\ref{EqDiffusionA}), is also not seriously taken into consideration} is employed for the analysis, we take the obtained values of $\kappa_{xx}$ to be acceptably close to the expected values $\kappa_{xx}^\mathrm{WF}$.
We stress that the procedure described here provides a handy method to roughly estimate the thermal conductivity using only a simple experimental setup for standard ac lock-in resistance measurement and a standard Hall bar device.

So far, we have employed Eq.\ (\ref{Eqthermalcond}) to deduce $\kappa_{xx}$ using material parameters taken from the literature and assuming the values of the relevant temperatures.
Conversely, one can, in principle, employ Eq.\ (\ref{Eqthermalcond}) to estimate $D$, $h_{14}$ and $T_\mathrm{d}$ by postulating that $\kappa_{xx}^\mathrm{WF}$ gives the correct value of the thermal conductivity. 
By first determining $T_d$ from lower temperature region and then by performing the fitting employing $D$ and $h_{14}$ as fitting parameters, we can let $\kappa_{xx}$ in Eq.\ (\ref{Eqthermalcond}) reproduce $\kappa_{xx}^\mathrm{WF}$ within $\sim$10\% by taking $T_d (\mathrm{K}) = (1/2) \{ T_e (\mathrm{K}) - 0.3222 - \sqrt{0.01384+[T_e (\mathrm{K}) - 0.3222]^2} \}+0.3222$, \footnote{This formula is less reliable at $T_e \agt 0.5$ K, where $T_d$ is obtained by extrapolating the low-temperature values since the little dependence of $\kappa_{xx}$ on $T_d$ hindered us from deducing $T_d$ directly} $h_{14} = 1.44 \times 10^9$ V/m, and $D = 0$ (namely, by neglecting the deformation-potential contribution). Note, however, that the reliability of these values is severely limited by the oversimplified model mentioned above.

\section{Conclusions}
We have shown that the Joule heating of a GaAs/AlGaAs 2DEG by an ac current with the angular frequency $\omega$, accompanied by the oscillating electron temperature $T_e$ having the angular frequency $2\omega$, can be monitored by detecting the fundamental ($\omega$) and the third-harmonic ($3\omega$) contents of the resulting voltage drop.
We have deduced the highest ($T_\mathrm{H}$) and the lowest ($T_\mathrm{L}$) temperatures of the oscillating $T_e$ for various values of the heating current. Employing the temperature response to the Joule heating thus acquired, we have further drawn out the thermal conductivity $\kappa_{xx}$ of the 2DEG, with the aid of the simple formula we have deduced in a recent publication \cite{Endo19} modeling the thermal flux through the voltage arms, and the well-known formulas for the power per area transferred from the 2DEG to the lattice system. \cite{Endo19,Price82,KajiokaCO13}
The resulting $\kappa_{xx}$ is found to be roughly in agreement with the thermal conductivity estimated from the electrical conductivity via the Wiedemann-Franz law. 
The procedure we have taken here to deduce $\kappa_{xx}$, corresponding to 2D analog of the 3$\omega$ method, provides us with a convenient method to roughly estimate the thermal conductivity using only a simple Hall-bar sample and the conventional experimental setup for the resistance measurement by the ac lock-in technique.

\section*{Supplementary Material}
See the Supplementary Material for I.\ $R_\omega$ and $R_{3\omega}$ taken with various frequencies at $I_0 =1$ $\mu$A, II.\  $R_\omega$ and $R_{3\omega}$ taken at various temperatures at $I_0 =10$ nA,  III.\ measurements and analyses similar to those shown in Sec.\  \ref{SecExp} of the main text but performed at an elevated bath temperature $T_\mathrm{bath} = 600$ mK, and IV.\ procedure for deducing the quantum mobility $\mu_q$.

\begin{acknowledgments}
This work was supported by JSPS KAKENHI Grants No. JP20K03817 and No. JP19H00652.
\end{acknowledgments}

\appendix*
\section{The dimensionless functions $G_s^r(n_e,T)$ \label{SecAppendix}}
In this Appendix, we present the detailed description of the dimensionless functions $G_s ^r (n_e,T)$ ($r = $ df, pz and $s = l, t$)  in Eq.\ (\ref{EqPi}) for completeness. The functions are written as \cite{Price82,KajiokaCO13}
\begin{eqnarray}
G_s ^r (n_e,T) \equiv &
\displaystyle{\frac{1}{\pi} \int_{-\infty}^\infty  d\zeta \left| F (q_{Ts} \zeta ) \right|^2 \times \hspace{30mm}} \nonumber \\
 & \! \! \! \! \! \! \! \! \! \! \! \! \displaystyle{\int_0^{\kappa_\text{F}} \! \! \! \! \! \frac{d\xi}{\sqrt{1 - (\xi / \kappa_\text{F})^2}} \frac{g_s^r (\xi , \zeta)}{e^{\sqrt{\xi ^2  + \zeta ^2 } }  - 1} \frac{1}{H^2( q_{Ts} \xi )} },
\label{Gsr}
\end{eqnarray}
with
\begin{equation}
F( q_z ) = \int {dz |\Phi (z)| ^2} e^{iq_z z},
\label{FormFactor}
\end{equation}
and
\begin{equation}
H( q_{\|} ) = \iint {dz_1 dz_2 |\Phi (z_1)| ^2  |\Phi (z_2)| ^2} e^{-q_{\|} |z_1 - z_2|},
\label{ScrnFactor}
\end{equation}
where $q_z$ and $q_\|$ are the components of the phonon wavevector perpendicular and parallel to the 2DEG plane, respectively, $\Phi (z)$ represents the envelope of the 2DEG wavefunction in the $z$ direction,  and $\kappa_\text{F} \equiv 2 k_\text{F}/q_{Ts}$, $\xi \equiv q_\|/q_{Ts}$, $\zeta \equiv q_z/q_{Ts}$ with $q_{Ts} \equiv k_\text{B} T / \hbar v_s$ representing the typical wavenumber of the acoustic phonons at the temperature $T$.
The kernels $g_s^r (\xi , \zeta)$ in Eq.\ (\ref{Gsr}) are given by \cite{Price82,KajiokaCO13}
\begin{equation}
g_l ^\text{def} (\xi , \zeta) \equiv \xi ^2 (\xi ^2 + \zeta ^2) ^{3/2},
\label{gdefl}
\end{equation}
\begin{equation}
g_l ^\text{pz} (\xi , \zeta) \equiv 
\frac{ 9 \xi ^6 \zeta ^2 }{2 (\xi ^2 + \zeta ^2) ^{5/2} },
\label{gpzl}
\end{equation}
and
\begin{equation}
g_t ^\text{pz} (\xi , \zeta) \equiv 
\frac{ 8 \xi ^4 \zeta ^4 + \xi ^8}{4 (\xi ^2 + \zeta ^2) ^{5/2} }.
\label{gpzt}
\end{equation}
Noting that $q_{Ts}$ is smaller than the inverse of the rms thickness $\sim$5 nm of our 2DEG \cite{Endo05MA} in the temperature range $T_\mathrm{p} < T_e \alt 1.2$ K encompassed in the present study, we can make an approximation $\Phi (z) \simeq \delta (z)$ and thus replace $|F(q_z)|$ and $H(q_{\|})$ in Eq.\ (\ref{Gsr}) by unity to good approximation.

\section*{AUTHOR DECLARATIONS}
\subsection*{Conflict of Interest}
The authors have no conflicts to disclose.

\section*{DATA AVAILABILITY}
The data that support the findings of this study are available from the corresponding author upon reasonable request.



%
%

%


\bibliography{ourpps,thermo,lsls,twodeg}

\end{document}



\title{{\footnotesize Supplementary Material} \\ 
Joule heating and the thermal conductivity of a two-dimensional electron gas at cryogenic temperatures studied by modified 3 $\omega$ method} 



\author{Akira Endo}
\email[]{akrendo@issp.u-tokyo.ac.jp}
\affiliation{The Institute for Solid State Physics, The University of Tokyo, 5-1-5 Kashiwanoha, Kashiwa, Chiba 277-8581, Japan}

\author{Shingo Katsumoto}
\affiliation{The Institute for Solid State Physics, The University of Tokyo, 5-1-5 Kashiwanoha, Kashiwa, Chiba 277-8581, Japan}

\author{Yasuhiro Iye}
\affiliation{The Institute for Solid State Physics, The University of Tokyo, 5-1-5 Kashiwanoha, Kashiwa, Chiba 277-8581, Japan}



\begin{abstract}
\end{abstract}

\maketitle 

\section{Frequency dependence of $R_\omega$ and $R_{3\omega}$}
\begin{figure}[p]
\includegraphics[width=8.4cm,clip]{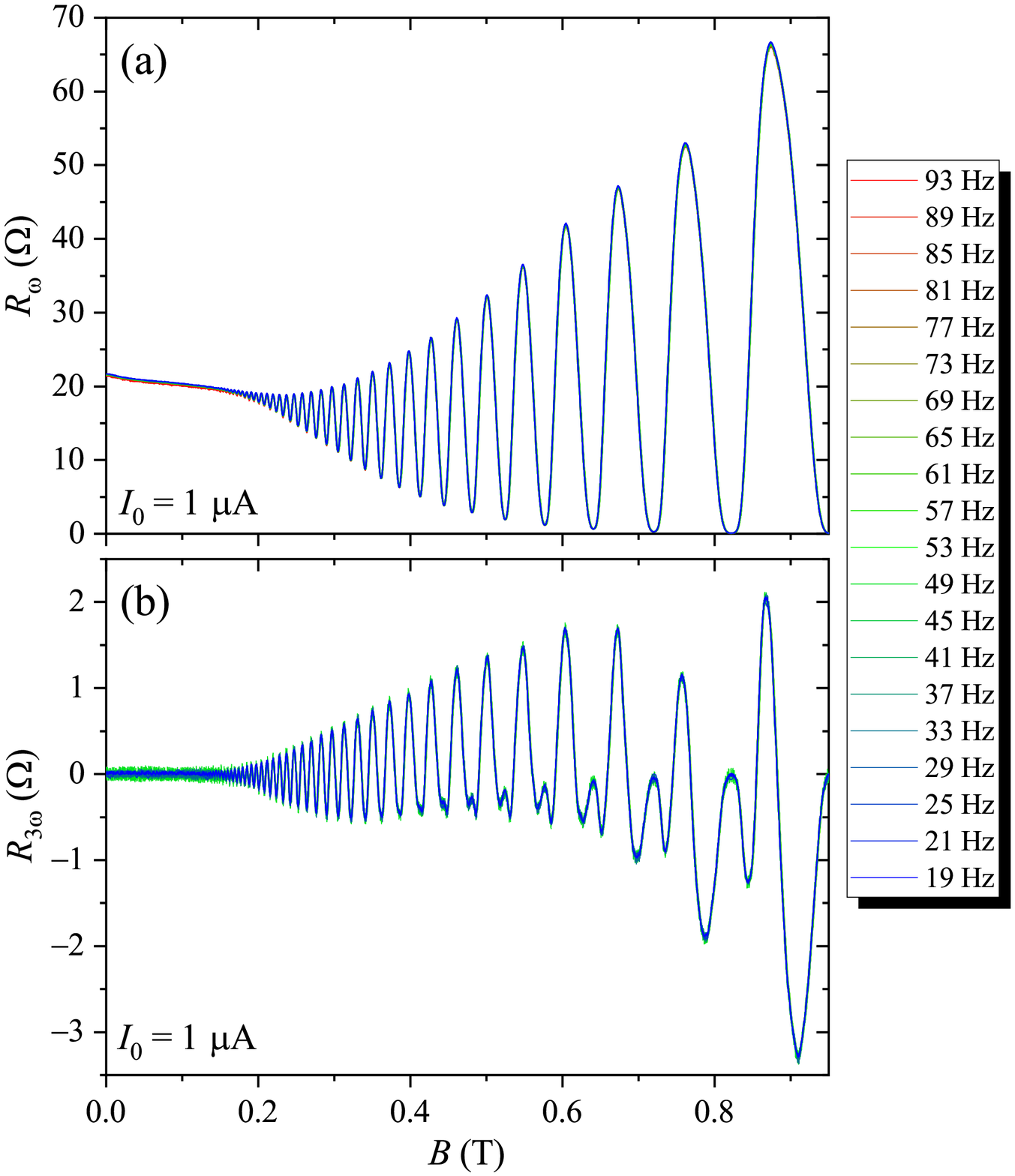}%
\caption{The $\omega$ (a) and $3 \omega$ (b) components of the resistance $R_{\omega}$ and $R_{3\omega}$ for various values of the frequency, ranging from 19 Hz to 93 Hz, of the ac heating current with $I_0 =1$ $\mu$A\@. 
\label{R1R3fdep}}
\end{figure}

We have performed the measurements of $R_\omega$ and $R_{3\omega}$ at frequencies $f = \omega / (2\pi)$ varying from 19 Hz to 93 Hz for all the heating current $I_0$ presented in the main text. Both $R_\omega$ and $R_{3\omega}$ were found to exhibit virtually no frequency dependence in this frequency range, as exemplified by Fig\ \ref{R1R3fdep} for $I_0 = 1$ $\mu$A (except for the slight deterioration of the signal-to-noise ratio at the frequencies close to the commercial line frequency 50 Hz). 
So far, we have limited ourselves to the real part (in-phase oscillations) of $R_\omega$ and $R_{3\omega}$, although the measurements have been performed simultaneously for the imaginary parts (out-of-phase oscillations) $R_{\omega,\mathrm{out}}$ and $R_{3\omega,\mathrm{out}}$ as well. Both $R_{\omega,\mathrm{out}}$ and $R_{3\omega,\mathrm{out}}$ showed slight frequency dependence. Unfortunately, however, the frequency dependence is mainly attributable to an artifact caused by parasitic capacitive coupling between the wires measuring the diagonal and the Hall resistances. \cite{Endo01c} In order to correctly investigate the frequency dependence, it will be necessary to replace the electrical wires in our dilution-fridge probe to coaxial cables so that the capacitive coupling can be eliminated. Coaxial cables will also be beneficial to extend the measurements up to orders of magnitude higher frequency ranges.

\clearpage
\section{Temperature dependence of $R_\omega$ and $R_{3\omega}$}
\begin{figure}[p]
\includegraphics[width=8.4cm,clip]{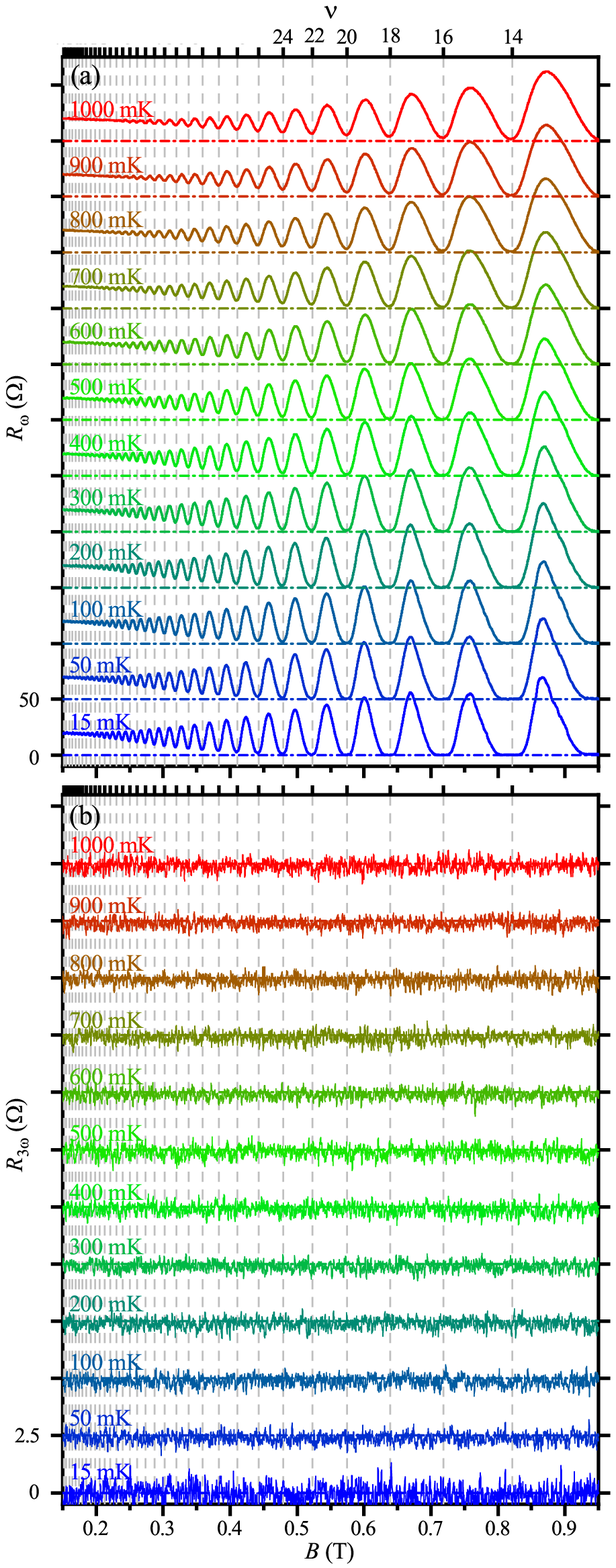}%
\caption{The $\omega$ (a) and $3 \omega$ (b) components of the resistance $R_{\omega}$ and $R_{3\omega}$ measured at various bath temperatures $T_\mathrm{bath}$ noted in the figure. Traces are sequentially offset by 50 $\Omega$ and 2.5 $\Omega$ in (a) and (b), respectively, for clarity, with increasing $T_\mathrm{bath}$. The vertical dashed lines and the horizontal dot-dashed lines indicate the positions of the even-integer fillings (top axis) and the zero for the corresponding trace, respectively. $I_0 = 10$ nA\@.
\label{R1R3Ts}}
\end{figure}

We have measured the dependence of $R_\omega$ and $R_{3\omega}$ on the bath temperature $T_\mathrm{bath} = 15$$-$$1000$ mK, employing a small current $I_0 = 10$  nA for which the heating effect is negligibly small. As shown in Fig.\ \ref{R1R3Ts} (a), $R_\omega$ exhibits usual SdHO with the amplitude decreasing with increasing $T_\mathrm{bath}$, bearing resemblance to the $I_0$ dependence shown in Fig.\ \R1R3Is (a) in the main text. The third-harmonic component $R_{3\omega}$, by contrast, does not show any discernible signal regardless of $T_\mathrm{bath}$ [Fig.\ \ref{R1R3Ts} (b)] due to the absence of measurable heating effect.

\clearpage
\section{Measurements at an elevated bath temperature}
\begin{figure}[H]
\includegraphics[width=8.4cm,clip]{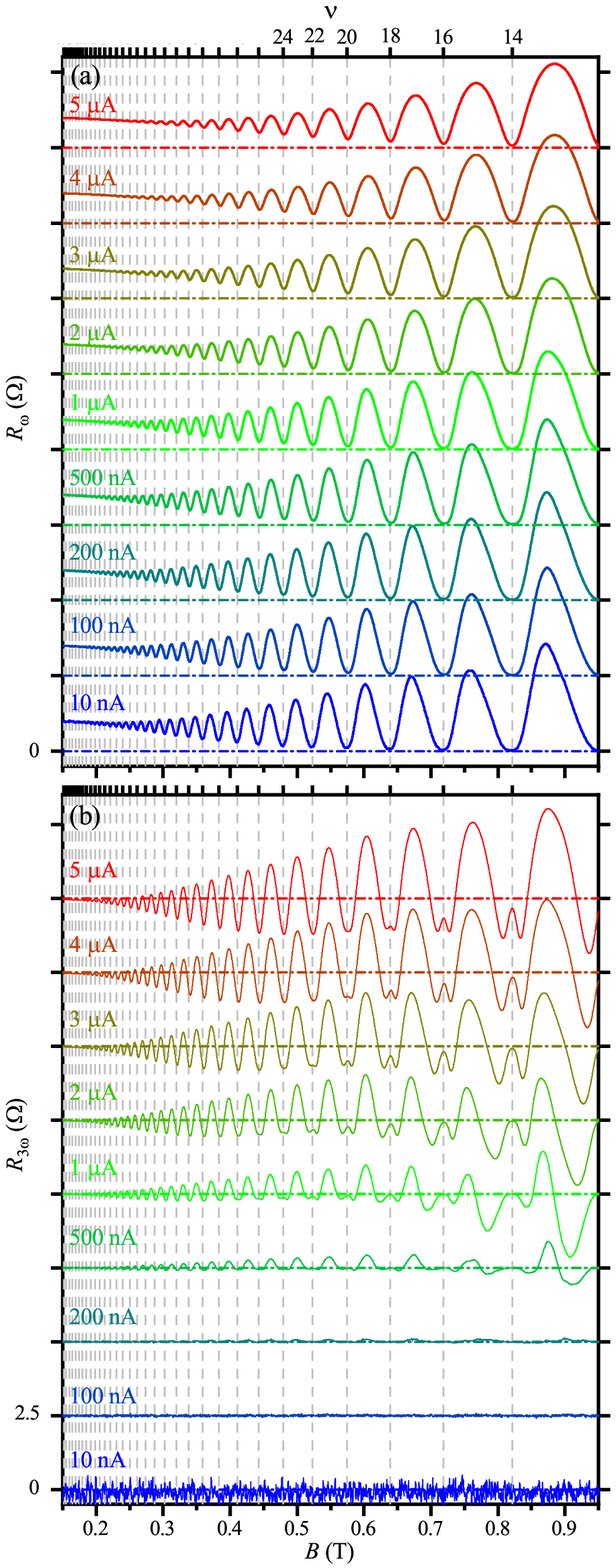}%
\caption{The $\omega$ (a) and $3 \omega$ (b) components of the resistance $R_{\omega}$ and $R_{3\omega}$ measured at $T_\mathrm{bath} = 600$ mK for various values of the heating current $I_0$ noted in the figures.  Traces are sequentially offset by 50 $\Omega$ and 2.5 $\Omega$ in (a) and (b), respectively, for clarity, with increasing $I_0$. The vertical dashed lines and the horizontal dot-dashed lines indicate the positions of the even-integer fillings (top axis) and the zero for the corresponding trace, respectively. 
\label{R1R3IsT06}}
\end{figure}

We have repeated the measurements and analyses of $R_{\omega}$ and $R_{3\omega}$ described in section \SecExp \ of the main text, with the bath temperature $T_\mathrm{bath} = 15$ mK (base temperature) replaced by an elevated temperature $T_\mathrm{bath} = 600$ mK\@. In Fig.\ \ref{R1R3IsT06} (a) and (b), we plot $R_{\omega}$ and $R_{3\omega}$, respectively, measured at $T_\mathrm{bath} = 600$ mK for various values of $I_0$ ranging from 10 nA to 5 $\mu$A\@. Comparison of Fig.\ \ref{R1R3IsT06} (b) with its counterpart at $T_\mathrm{bath} = 15$ mK [Fig.\ \R1R3Is (b) in the main text] reveals that the oscillation signals in $R_{3\omega}$ becomes smaller at the higher $T_\mathrm{bath}$, especially at lower $I_0$. 
This is mainly because the Joule heating by a smaller $I_0$ does not introduce enough temperature oscillations to be detected by $R_{3\omega}$ at the higher $T_\mathrm{bath}$.

\begin{figure}
\includegraphics[width=8.4cm,clip]{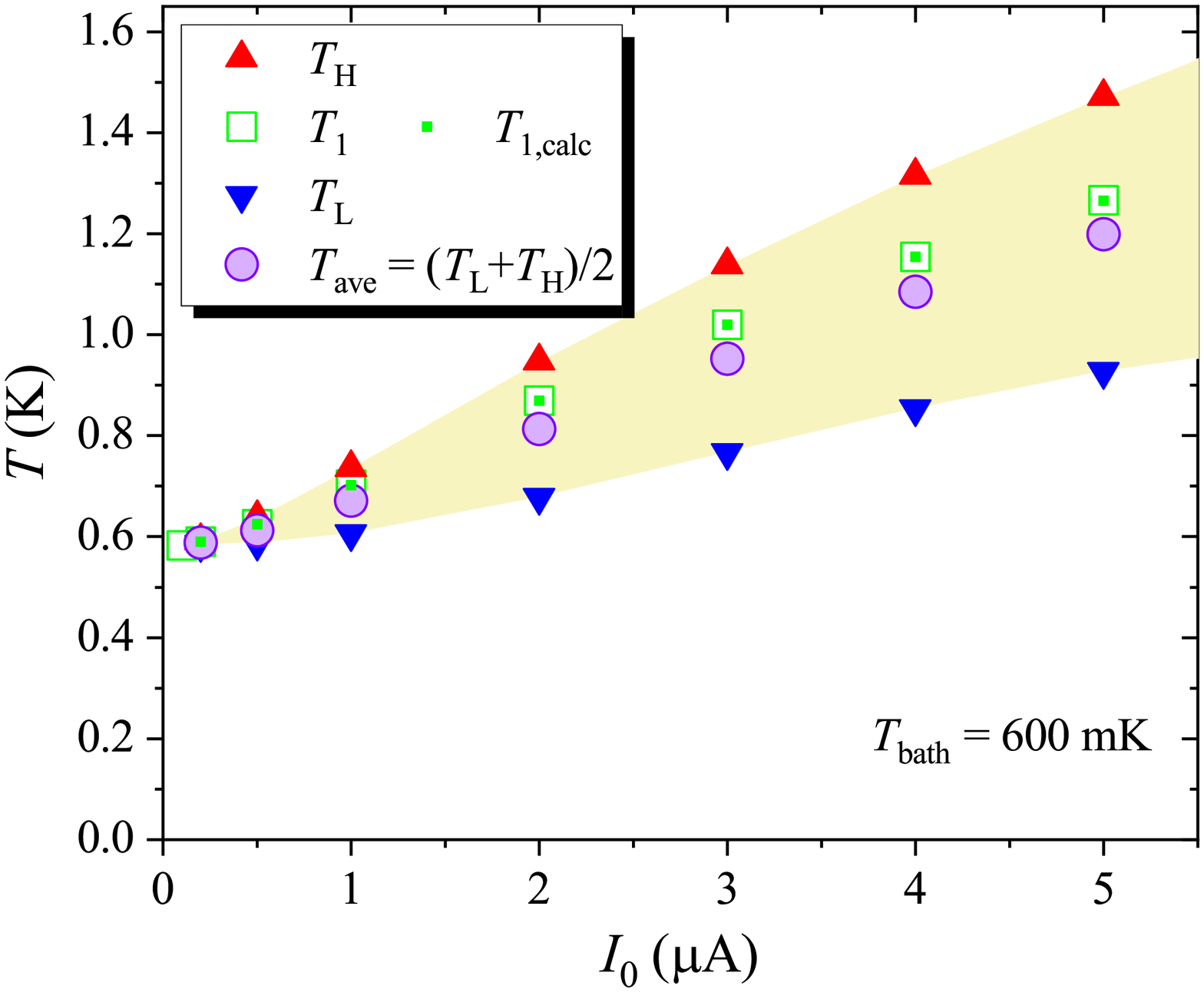}%
\caption{$T_\mathrm{L}$ and $T_\mathrm{H}$ at $T_\mathrm{bath} = 600$ mK for various values of $I_0$, obtained from $R_\omega$ and $R_{3\omega}$ shown in Fig.\ \ref{R1R3IsT06} following the procedure described in the main text. The average temperature, $T_\mathrm{ave} = (T_\mathrm{L}+T_\mathrm{H})/2$, the temperature $T_1$ obtained from the analysis of the SdHOs in $R_\omega$, and $T_{1,\mathrm{calc}}$ calculated by Eq.\ (\T1calc) in the main text are also plotted.
\label{I0vsT06K}}
\end{figure}

Following the analyses described in the main text, we calculated the highest ($T_\mathrm{H}$) and the lowest ($T_\mathrm{L}$) temperatures during a cycle of the temperature oscillations using $R_\omega$ and $R_{3\omega}$ shown in Fig.\ \ref{R1R3IsT06}. These temperatures are plotted as a function of $I_0$ in Fig.\ \ref{I0vsT06K}, along with the average temperature $T_\mathrm{ave}$, the temperature $T_1$ obtained from the conventional analysis of the SdHOs in $R_\omega$, and $T_{1,\mathrm{calc}}$ calculated from $T_\mathrm{H}$ and $T_\mathrm{L}$ using Eq.\ (\T1calc) in the main text.
We can see that the temperature increment $\Delta T = T_\mathrm{H} - T_\mathrm{L}$ at $T_\mathrm{bath} = 600$ mK is smaller, at low $I_0$, than its counterpart at the base temperature shown in Fig.\ \TLTHderiv (d) in the main text, but gradually approaches the low-$T_\mathrm{bath}$ value with increasing $I_0$.

\begin{figure}[H]
\includegraphics[width=8.4cm,clip]{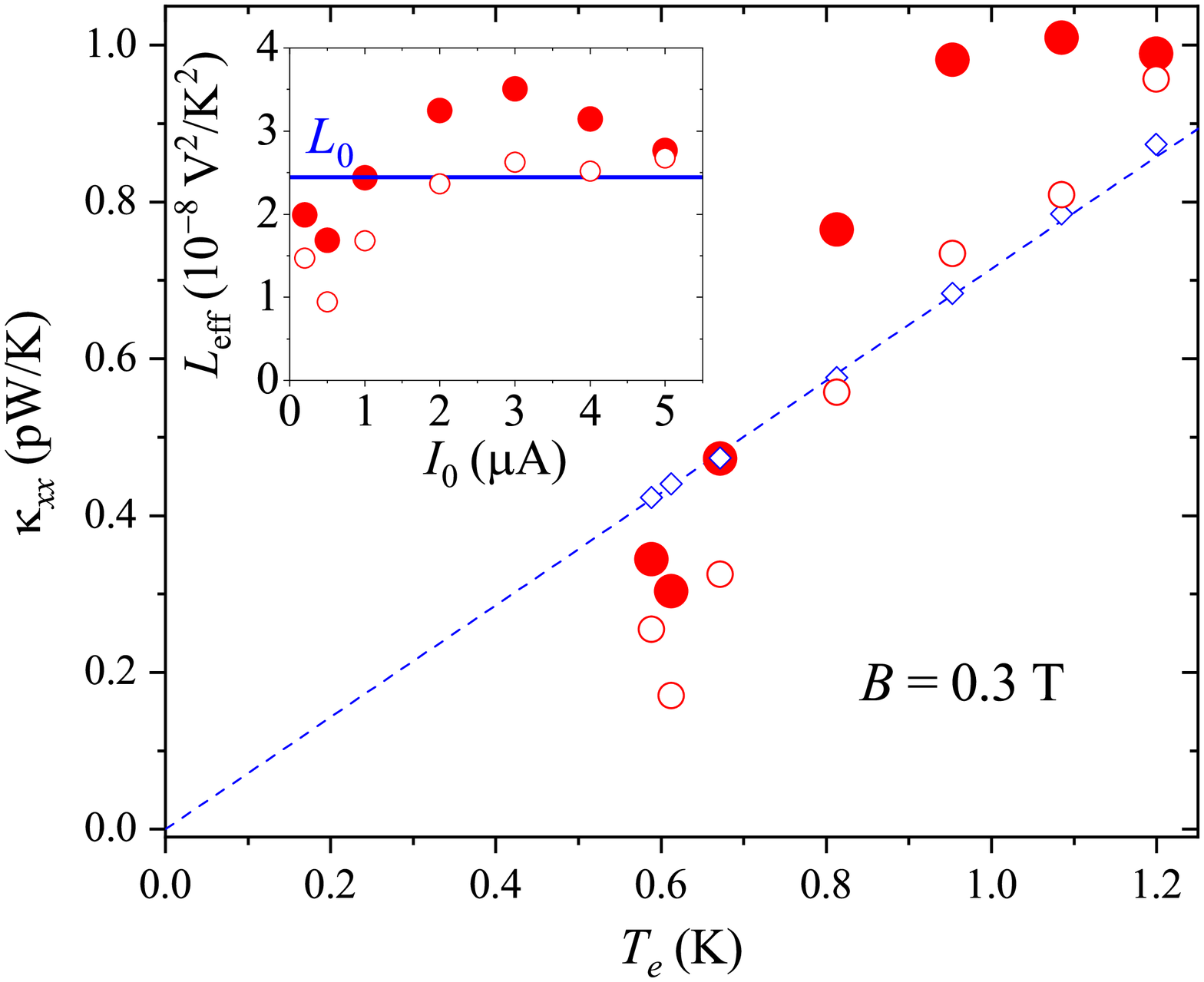}%
\caption{Thermal conductivity $\kappa_{xx}$ calculated by Eq.\ (\Eqthermalcond) in the main text with $T_e = T_\mathrm{ave}$ at $B = 0.3$ T, using the parameters $D = -8.33$ eV and $h_{14} = 1.2\times 10^9$ V/m (solid circles) or $D = 0$ and $h_{14} = 1.44\times 10^9$ V/m (open circles). The values of $\kappa_{xx}^\mathrm{WF}$ calculated by the Wiedemann-Franz law, Eq.\ (\EqWF) in the main text, is plotted by open diamonds. (Dashed line is an eye guide). Inset: effective Lorenz number $L_\mathrm{eff}$ given by Eq.\ (\Lorenz) in the main text plotted against $I_0$. The horizontal line indicates the value $L_0$ corresponding to the Wiedemann-Franz law. 
\label{thermcondT06}}
\end{figure}

The thermal conductivities $\kappa_{xx}$ calculated with the temperatures in Fig.\ \ref{I0vsT06K} using Eq.\ (\Eqthermalcond) in the main text, assuming $T_e = T_\mathrm{ave}$ and $T_\mathrm{d} =T_\mathrm{bath}$, are plotted in Fig.\ \ref{thermcondT06} to be compared with Fig.\ \thermalcond \  in the main text. The values of $\kappa_{xx}$ calculated using the sample parameters taken from the literature noted in the main text are plotted by solid circles. Deviation at a higher $T_e$ from the thermal conductivity $\kappa_{xx}^\mathrm{WF}$ calculated by the Wiedemann-Franz law becomes more prominent for the higher $T_\mathrm{bath}$. We presume this to be mainly resulting from the uncertainty in the literature values of the electron-phonon coupling parameters $D$ and $h_\mathrm{14}$.   
Replacing the values of $D$ and $h_{14}$ with those obtained by the fitting at $T_\mathrm{bath} = 15$ mK deduced in the main text ($D = 0$ and $h_{14} = 1.44\times 10^9$ V/m), $\kappa_{xx}$, plotted with the open circles in Fig.\ \ref{thermcondT06}, can be made closer to $\kappa_{xx}^\mathrm{WF}$.

\bibliography{ourpps,thermo,lsls,twodeg}

\section{Determining the quantum mobility $\mu_q$}
\begin{figure}[H]
\includegraphics[width=8.4cm,clip]{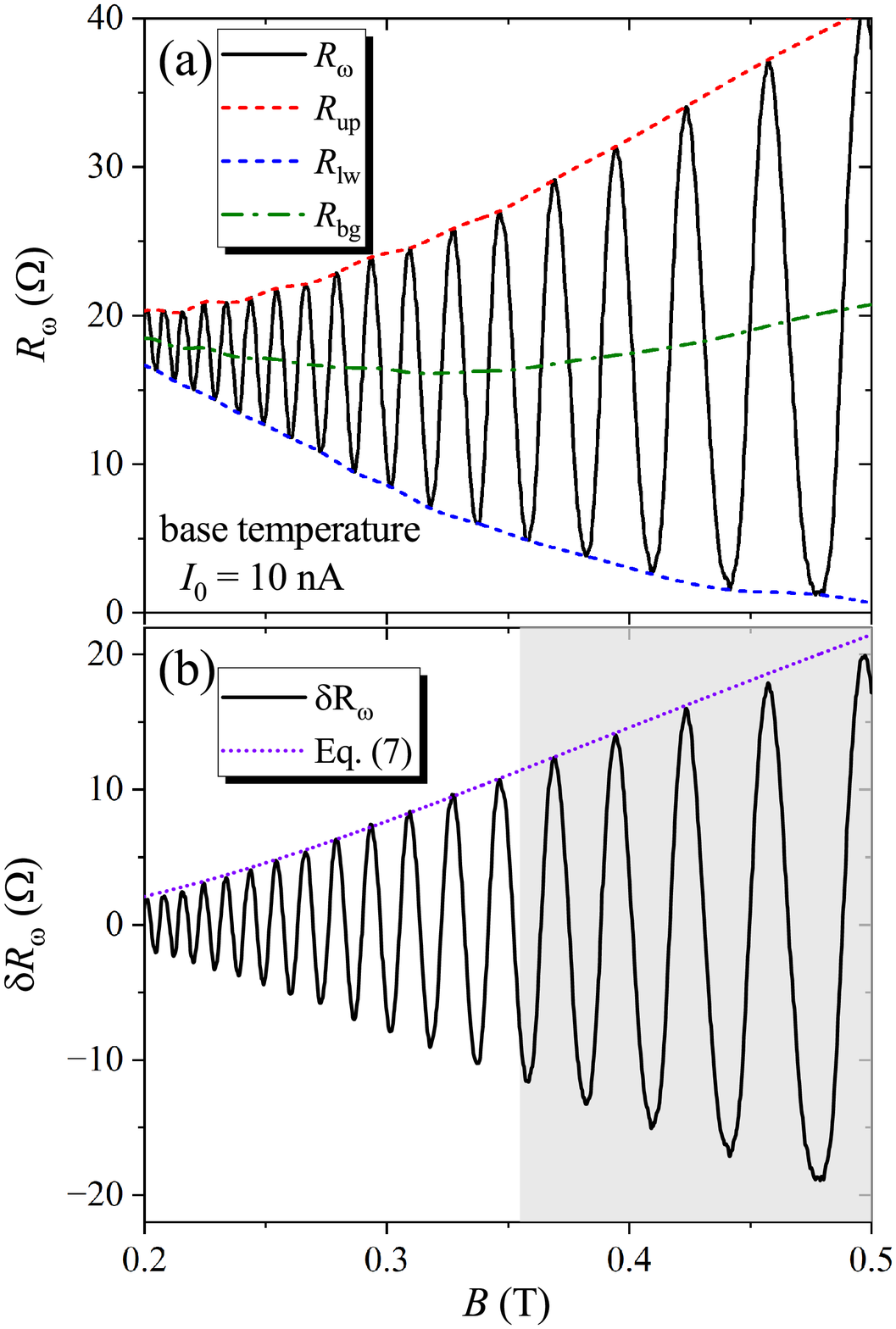}%
\caption{(a) $R_\omega$ taken with $I_0 = 10$ nA at the base temperature [a part of the lowermost trace in Fig.\ \R1R3Is (a) in the main text, or in Fig.\ \ref{R1R3Ts} (a)]  (solid line), upper ($R_\mathrm{up}$) and lower ($R_\mathrm{lw}$) envelops (dashed lines), and the slowly-varying background $R_\mathrm{bg}$ (dot-dashed line). (b) The oscillatory part $\delta R_\omega = R_\omega - R_\mathrm{bg}$ (solid line) and the fitting of the amplitude to Eq.\ (\EqSdHamp) in the main text (dotted line).  \label{SdH10nA}}
\end{figure}

As described in \SSecTe of the main text,  we extract the quantum mobility $\mu_ q$ of our 2DEG from the decay of the SdH oscillations taken with the small current $I_0 = 10$ nA at the base temperature $T_\mathrm{bath} = 15$ mK\@. The process is outlined in Fig.\ \ref{SdH10nA}. As delineated in the main text, we first deduce the slowly varying background $R_\mathrm{bg}$ as an average of the upper ($R_\mathrm{up}$, red dashed line) and the lower ($R_\mathrm{lw}$, blue dashed line) envelops [Fig.\ \ref{SdH10nA} (a)]. We then fit [Fig.\ \ref{SdH10nA} (b)] the amplitude of the oscillatory part $\delta R_\omega$, obtained by subtracting the background, to Eq.\ (\EqSdHamp), assuming $T_e = T_\mathrm{bath}$, in the magnetic-field range $B \leq 0.35$ T, where the effect of the spin-splitting does not manifest itself in the resistance. We obtain $\mu_q = 4.1$ m$^2$/(Vs) as the result of the fitting.
 \vspace{50mm}



%
%

%

